\documentclass[aps,preprint,amssymb]{revtex4}

\usepackage{epsfig}

\def\pot{_{\rm pot}}

\def\xc{_{\rm xc}}

\def\rv{{\bf r}}

\def\fv{{\bf f}}

\def\Rv{{\bf R}}

\def\beq{\begin{equation}}
\def\eeq{\end{equation}}

\def\nf{N_e}

\begin{document}
\renewcommand{\thefootnote}{\fnsymbol{footnote}} 
\renewcommand{\theequation}{\arabic{section}.\arabic{equation}}

\title{Intracule densities in the
strong-interaction limit of density functional theory}

\author{Paola Gori-Giorgi,$^a$ Michael Seidl$^b$ and Andreas Savin$^a$}
\email[]{E-mail: paola.gori-giorgi@lct.jussieu.fr}

\affiliation{$^a$ Laboratoire de Chimie Th\'eorique, CNRS, Universit\'e Pierre et Marie Curie, 4 Place Jussieu, F-75252 Paris, France\\
$^b$ Institute of Theoretical Physics, University of Regensburg, D-93040 Regensburg, Germany}

\date{\today}

\begin{abstract}
\noindent The correlation energy in density functional theory can be expressed exactly in terms of the change in the probability of finding two electrons at a given distance $r_{12}$ (intracule density) when the electron-electron interaction is multiplied by a real parameter $\lambda$ varying between 0 (Kohn-Sham system) and 1 (physical system). In this process, usually called adiabatic connection, the one-electron density is (ideally) kept fixed by a suitable local one-body potential. While an accurate intracule density of the physical system can only be obtained from expensive wavefunction-based calculations, being able to construct good models starting from Kohn-Sham ingredients would highly improve the accuracy of density functional calculations. To this purpose, we investigate the intracule density in the $\lambda\to\infty$ limit of the adiabatic connection. This strong-interaction limit of density functional theory turns out to be, like the opposite non-interacting Kohn-Sham limit, mathematically simple and can be entirely constructed from the knowledge of the one-electron density. We develop here the theoretical framework and, using accurate correlated one-electron densities, we calculate the intracule densities in the strong interaction limit for few atoms. Comparison of our results with the corresponding Kohn-Sham and physical quantities provides useful hints for building approximate intracule densities along  the adiabatic connection of density functional theory.
\end{abstract}

\maketitle

\section{Introduction}
\label{intro}
Kohn-Sham (KS) density functional theory (DFT) (see, e.g., \cite{Koh-RMP-99}) is a successful method for electronic structure calculations, thanks to its unique combination of low computational cost and reasonable accuracy. In the Kohn-Sham formalism, the total energy of a many-electron system in the external potential $\hat{V}_{ne}=\sum_i v_{ne}(\rv_i)$ is rewritten as a functional of the one-electron density $\rho(\rv)$,
\beq
E[\rho]=T_s[\rho]+U[\rho]+E\xc[\rho]+\int d\rv\,v_{ne}(\rv)\,\rho(\rv).
\label{eq_Erho}
\eeq
In Eq.~(\ref{eq_Erho}), $T_s[\rho]$ is the kinetic energy of a non-interacting system of fermions (usually called KS system) having the same one-electron density $\rho$ of the physical, interacting, system. The Hartree energy $U[\rho]$ is the classical repulsion energy, $U[\rho]=\frac{1}{2}\int d\rv\int d\rv'\rho(\rv)\rho(\rv')|\rv-\rv'|^{-1}$, and the exchange-correlation functional $E\xc[\rho]$ must be approximated. 

Despite its success in scientific areas ranging from material science to biology, DFT is far from being perfect, and a huge effort is put nowadays in trying to improve the approximations for $E\xc[\rho]$ (for recent reviews see, e.g., \cite{Mat-SCI-02,PerRuzTaoStaScuCso-JCP-05}). The focus of a large part of the scientific community working in this area has shifted from seeking explicit functionals of the density (like the generalized gradient approximations - GGA), to implicit functionals, typically using the exact exchange $E_{\rm x}[\rho]$, which is only explicitly known in terms of the Kohn-Sham orbitals $\phi_i(\rv)$ (for a recent review, see \cite{KumKro-RMP-08}). In this framework, although DFT was originally formulated as a method ``without wavefunction'', it might be natural to go back and actually think of DFT approximations in terms of model wavefunctions. This way of thinking can be very helpful for building approximations, and for combining DFT with other many-body methods (see, e.g., \cite{LeiStoWerSav-CPL-97,IikTsuYanHir-JCP-01,KamTsuHir-JCP-02,PolSavLeiSto-JCP-02,TouColSav-PRA-04,AngGerSavTou-PRA-05,TouGorSav-TCA-05,GolWerSto-PCCP-05,GolWerStoLeiGorSav-CP-06,GerAngMarKre-JCP-07}).

The adiabatic connection formalism (for a review, see \cite{SavColPol-IJQC-03}) is a useful tool to think of DFT functionals in terms of wavefunctions. In its simpler and original version \cite{HarJon-JPF-74,LanPer-SSC-75,GunLun-PRB-76}, the electron-electron repulsion operator $\hat{V}_{ee}$ in the $N$-electron hamiltonian $\hat{H}$ (in Hartree atomic units used throughout),
\beq
\hat{H}=\hat{T}+\hat{V}_{ee}+\hat{V}_{ne},\qquad \hat{T}=-\sum_{i=1}^N \frac{\nabla^2_{\rv_i}}{2},\qquad \hat{V}_{ee}=\sum_{i>j=1}^N \frac{1}{|\rv_i-\rv_j|}, \qquad \hat{V}_{ne}=\sum_{i=1}^N v_{ne}(\rv_i),
\label{eq_Hphys}
\eeq
is multiplied by a real parameter $\lambda$, which varies between 0 and 1. At the same time, the external potential $v_{ne}(\rv)$ is replaced by another local potential, $v^\lambda(\rv)$, determined by the condition (allowed by the Hohenberg-Kohn theorems \cite{HohKoh-PR-64}, if $\rho$ is $v$-representable for all $\lambda$) that the one-electron density $\rho(\rv)$ do not change with $\lambda$. In this way, we define a set of hamiltonians $\hat{H}^\lambda$,
\beq
\hat{H}^\lambda=\hat{T}+\lambda\hat{V}_{ee}+\hat{V}^\lambda, \qquad \rho^\lambda(\rv)=\rho(\rv)\;\forall\,\lambda
\label{eq_Hlambda}
\eeq
all having the same $\rho(\rv)$ as the one of the physical hamiltonian of Eq.~(\ref{eq_Hphys}). In particular, at $\lambda=0$ we have the KS hamiltonian, i.e., the hamiltonian of a non-interacting system of fermions with the same density of the physical system, and $v^{\lambda=0}(\rv)=v_{\rm KS}(\rv)$, the familiar Kohn-Sham potential. If we denote by $\Psi^\lambda$ the ground-state wavefunctions of each hamiltonian $\hat{H}^\lambda$ of Eq.~(\ref{eq_Hlambda}), we easily find
\begin{eqnarray}
T_s[\rho] & = & \langle \Psi^{\lambda=0}|\hat{T}|\Psi^{\lambda=0}\rangle \\
E_{\rm x}[\rho] & = & \langle \Psi^{\lambda=0}|\hat{V}_{ee}|\Psi^{\lambda=0}\rangle -U[\rho] \label{eq_defEx} \\
E_c[\rho] & = & \int_0^1 \langle \Psi^{\lambda}|\hat{V}_{ee}|\Psi^{\lambda}\rangle d\lambda -E_{\rm x}[\rho]-U[\rho],
\label{eq_defEc}
\end{eqnarray}
where $\Psi^{\lambda=0}$ is, in most cases, a single Slater determinant formed by the KS orbitals $\phi_i$.

Equations (\ref{eq_defEx}) and (\ref{eq_defEc}) can be rewritten in terms of the intracule density $I(r_{12})$ (also called in the DFT community spherically and system-averaged pair density), which was first introduced in the historical paper of Coulson and Neilson \cite{CouNei-PPSL-61}. Since then, $I(r_{12})$ has been used by several authors to understand electronic correlation both in density functional theory (see, e.g., \cite{GunLun-PRB-76,BurPerErn-JCP-98,BurPer-IJQC-95}) and in post-Hartree-Fock methods (see, e.g., \cite{LesKra-JCP-66,Koh-JCP-72,Kat-PRA-72,RegTha-JPB-84,ThaTriSmi-IJQC-84,SarDomAguUga-JCP-92,CioLiu-JCP-98,ValUgaBoy-INC-00}). Given an $N$-electron wavefunction $\Psi$, the intracule density $I(r_{12})$ is defined as the integral of $|\Psi|^2$ over all variables but $r_{12}=|\rv_1-\rv_2|$,
\beq
I(r_{12}) = \frac{N(N-1)}{2}\sum_{\sigma_1...\sigma_N}
 \int |\Psi(\rv_{12},\Rv,\rv_3,...,\rv_N)|^2
\frac{d\Omega_{\rv_{12}}}{4\pi} d\Rv d\rv_3...d\rv_N,
\label{eq_defI}
\eeq
where $\rv_{12}=\rv_1-\rv_2$, and $\Rv=\frac{1}{2}(\rv_1+\rv_2)$. Here we have normalized $I(r_{12})$ to the number of electron pairs. The quantity $I(r_{12}) 4\pi r_{12}^2$ is proportional to the probability distribution for the electron-electron distance in the state described by the wavefunction $\Psi$. Gill and coworkers \cite{GilONe-JCP-05,GilCriONeBes-PCCP-06,DumCriGil-PCCP-07,CriDumGil-JCP-07} have defined an interesting ``family of intracules'', and made the hypothesis that the correlation energy of Hartree-Fock theory can be approximated as a linear functional of one of these intracules.

In terms of the intracule densities $I^\lambda(r_{12})$ associated to each wavefunction $\Psi^\lambda$ of the  adiabatic connection of Eq.~(\ref{eq_Hlambda}), the Kohn-Sham correlation energy $E_c[\rho]$ of Eq.~(\ref{eq_defEc}) can be rewritten exactly as
\beq
E_c[\rho]=\int_0^1 d\lambda \int d \rv_{12} \frac{I^\lambda(r_{12})-I^{\lambda=0}(r_{12})}{r_{12}}=\int_0^1 d\lambda \int_0^\infty d r_{12} \,4\pi \,r_{12}\, \left[I^\lambda(r_{12})-I^{\lambda=0}(r_{12})\right].
\label{eq_EcfromI}
\eeq 
Correlation in Kohn-Sham DFT is thus fully determined by the change in the intracule density when the electron-electron interaction is turned on  with the one-electron density $\rho(\rv)$ fixed. The difference $I^{\lambda=1}(r_{12})-I^{\lambda=0}(r_{12})$ determines the correction due to correlation to the expectation of $\hat{V}_{ee}$, and the integration over $\lambda$ recovers the correction to the expectation of $\hat{T}$. By construction, there is no correction to the expectation of $\hat{V}_{ne}$.

Starting from the observation that $I(r_{12})$ couples to the operator $\hat{V}_{ee}$ in the same way as $\rho(\rv)$ couples to $\hat{V}_{ne}$, i.e., that  the expectations $\langle\Psi|\hat{V}_{ne}|\Psi\rangle$ and $\langle\Psi|\hat{V}_{ee}|\Psi\rangle$ are given by linear functionals of $\rho(\rv)$ and $I(r_{12})$, respectively,  
\beq
\langle\Psi|\hat{V}_{ne}|\Psi\rangle=\int d\rv\,v_{ne}(\rv)\,\rho(\rv), \qquad
\langle\Psi|\hat{V}_{ee}|\Psi\rangle=\int d\rv_{12}\,\frac{1}{r_{12}}\,I(r_{12}),
\eeq
it is possible to derive an exact formalism \cite{GorSav-PRA-05,GorSav-PM-06,Nag-JCP-06,GorSav-IJMPB-07,GorSav-JCTC-07} in which a set of effective equations for each $I^\lambda(r_{12})$ along the DFT adiabatic connection is coupled to the KS equations to generate the correlation energy from Eq.~(\ref{eq_EcfromI}). In this computational scheme one needs to make two approximations: 
\begin{enumerate}
\item  the exact equation for $I^\lambda(r_{12})$ involves the solution of a many-body problem for a cluster of interacting fermions \cite{GorSav-PM-06}, which is approximated with a radial Schr\"odinger equation for $\sqrt{I^\lambda(r_{12})}$ \cite{Nag-JCP-06}, possibly divided into effective geminals $g_i^\lambda(r_{12})$ \cite{GorPer-PRB-01,GorSav-PRA-05,GorSav-IJMPB-07},
\beq
\left[-\nabla^2_{r_{12}}+w_{\rm eff}^\lambda(r_{12})\right]\,g_i^\lambda(r_{12})=\epsilon_i^\lambda\,g_i^\lambda(r_{12}), \qquad I^\lambda(r_{12})=\sum_{i=1}^{N_g} \nu_i\,|g_i^\lambda(r_{12})|^2,
\label{eq_eqsforI}
\eeq
for which one needs to choose the number $N_g$ and the occupancy $\nu_i$; 
\item an approximation for $w_{\rm eff}^\lambda(r_{12})$ needs to be designed.
\end{enumerate} 
As far as point 1 is concerned, we can say that the choice $N_g=1$ is always possible \cite{Nag-JCP-06}, and yields good results in the uniform electron gas when combined with an approximation for $w_{\rm eff}^\lambda(r_{12})$ inspired to the Fermi-hypernetted-chain approach \cite{DavAsgPolTos-PRB-03}. Again in the case of the uniform electron gas, the choice of a determinant-like occupancy for the effective geminals ($N_g=N(N-1)/2$, $\nu_i=1$ (3) for even (odd) relative angular momentum states) yields accurate results with much simpler approximations for $w^\lambda_{\rm eff}(r_{12})$ \cite{GorPer-PRB-01,DavPolAsgTos-PRB-02}. In general, the choice of using localized geminals would make it easier to impose size consistency. 

Regarding point 2, the basic idea is to write $w_{\rm eff}^\lambda(r_{12})$  as
\beq
w_{\rm eff}^\lambda(r_{12})=w_{\rm eff}^{\lambda=0}(r_{12})+\frac{\lambda}{r_{12}}
+w_c^\lambda(r_{12}).
\label{eq_wc}
\eeq
 The interaction $w_{\rm eff}^{\lambda=0}(r_{12})$ is the one that, when inserted into Eqs.~(\ref{eq_eqsforI}), yields the intracule density of the KS system, $I^{\lambda=0}(r_{12})$, which can be constructed by inserting the KS Slater determinant into Eq.~(\ref{eq_defI}). In this step, the analytical integrals developed by Gill and coworkers \cite{GilONe-JCP-05,GilCriONeBes-PCCP-06,DumCriGil-PCCP-07,CriDumGil-JCP-07} to calculate Hartree-Fock intracules may turn extremely useful. In Eq.~(\ref{eq_wc}) the term $\lambda/r_{12}$ ensures that the corresponding $I^\lambda(r_{12})$ satisfies the electron-electron cusp condition (see, e.g., \cite{RajKimBan-PRB-78}). We need then to approximate $w_c^\lambda(r_{12})$, an effective potential that should essentially ``tell'' to the intracule density that, while the electron-electron interaction is turned on, the one-electron density $\rho(\rv)$ does not change. As the information on $\rho(\rv)$ has been ``washed away'' in the integration over the center of mass $\Rv$ of Eq.~(\ref{eq_defI}), this constraint can be imposed only in an approximate way. For two-electron systems, for which Eq.~(\ref{eq_eqsforI}) is exact with one geminal, simple approximations (based on the same ideas used in the uniform electron gas) for  $w_c^\lambda(r_{12})$ give accurate results
 \cite{GorSav-PRA-05,GorSav-IJMPB-07,GorSav-JCTC-07}. 

To go one step further, that is being able to construct approximations that work for many-electron systems of nonuniform density, a crucial issue is to investigate the effect on $I^\lambda(r_{12})$ of the constraint of fixed  $\rho(\rv)$ as $\lambda$ increases. To this purpose, in this paper we address the following question: what happens to $I^\lambda(r_{12})$ when $\lambda\to\infty$ ? Although at first glance this question may seem purely academic, there are several reasons for investigating this strong-interaction limit of DFT. The intracule density of the physical system ($\lambda=1$) can be obtained only from expensive wavefunction-based calculations (see, e.g., \cite{GalBueSar-CPL-03,Tha-CPL-03,TouAssUmr-JCP-07} and references therein), while in the $\lambda\to\infty$ limit the many-electron problem becomes mathematically simple, and we have recently shown  \cite{SeiGorSav-PRA-07} that a solution can be constructed starting from the density $\rho(\rv)$ only. The $\lambda\to\infty$ limit tells us what is the maximum extent to which the electrons can avoid each other without breaking the constraint of being in the given density $\rho(\rv)$. This information can be very useful for constructing approximations. Last but not least, the strong interaction limit can be used to build interpolations between the non-interacting KS limit ($\lambda=0$) and the $\lambda\to\infty$ limit, yielding an approximation for the physical ($\lambda=1$) system \cite{SeiPerKur-PRL-00}. 

This paper is organized as follows. In Sec.~\ref{sec_theory} we derive and discuss the equations needed to calculate the intracule density in the strong-interaction limit ($\lambda\to\infty$) of DFT. In Sec.~\ref{sec_app}, we apply the equations of Sec.~\ref{sec_theory} to calculate the $\lambda\to\infty$ intracule densities of small atoms, by using accurate correlated one-electron densities $\rho(\rv)$ as input. The results are then analyzed and discussed in Sec.~\ref{sec_disc}, and the last Sec.~\ref{sec_conc} is devoted to conclusions and perspectives.

\section{Theory}
\label{sec_theory}
The strong-interaction limit of DFT is defined by the $\lambda\to\infty$ limit of the hamiltonians of Eq.~(\ref{eq_Hlambda}) \cite{SeiPerLev-PRA-99,Sei-PRA-99,SeiGorSav-PRA-07}. The mathematical details of this limit can be found in \cite{SeiGorSav-PRA-07}. Here, we  briefly summarize the physical ideas that lie behind the theory, only reporting the equations that will be used in the following sections.
 
As $\lambda$ grows, the electrons repel each other more and more strongly. However, they are forced by the external potential $\hat{V}^\lambda$ of Eq.~(\ref{eq_Hlambda}) to yield the density $\rho(\rv)$. As $\lambda\to\infty$, it can be shown \cite{SeiPerLev-PRA-99,Sei-PRA-99,SeiGorSav-PRA-07} that, in order to keep the electrons in the density $\rho(\rv)$,  $\hat{V}^\lambda$  must be proportional to $\lambda$, $\hat{V}^{\lambda\to\infty}\to\lambda \hat{V}$. In this limit, the kinetic energy becomes negligible (of orders $\sqrt{\lambda}$ \cite{Sei-PRA-99}), and the solution of $\hat{H}^{\lambda\to\infty}$ reduces to a classical equilibrium problem for the 3$N$ dimensional function 
\beq
E_{\rm pot}(\rv_1,...,\rv_N)=\sum_{i>j=1}^N\frac{1}{|\rv_i-\rv_j|}+\sum_{i=1}^N v(\rv_i), \qquad v(\rv)=\lim_{\lambda\to\infty} \frac{v^\lambda(\rv)}{\lambda}. 
\label{eq_Epot}
\eeq
The square of the corresponding wavefunction, $|\Psi^{\lambda\to\infty}|^2$, becomes a distribution that is zero everywhere except in the configurations $(\rv_1^{(0)},...\rv_N^{(0)})$ for which  $E_{\rm pot}(\rv_1,...,\rv_N)$ has its absolute minimum. Typically, for a reasonable attractive potential $v(\rv)$, $E_{\rm pot}$ has a discrete set of minimizing configurations. In this case, however, the density corresponding to $|\Psi^{\lambda\to\infty}|^2$ would be given by a sum of peaks centered in the minimizing positions $\rv_i^{(0)}$, $\rho(\rv)\propto \sum_i \delta(\rv-\rv_i^{(0)})$. In order to get a smooth density like the one we find in the quantum mechanical problem at $\lambda=1$, we need a special potential $v(\rv)$ in Eq.~(\ref{eq_Epot}): the potential $v(\rv)$ must make the minimum of the 3$N$-dimensional function $E_{\rm pot}$ degenerate over the 3-dimensional subspace $M$ \cite{SeiGorSav-PRA-07}
\beq
M=\{\rv_1=\rv,\quad \rv_2=\fv_2(\rv),\quad \dots,\quad \rv_N=\fv_N(\rv),\quad \rv\in P\},
\eeq   
where $P$ is the region of space in which $\rho(\rv)\neq 0$. From the physical point of view, the distribution $|\Psi^{\lambda\to\infty}|^2$, which is zero everywhere except on $M$, describes a state in which the position of one of the electrons can be freely chosen in $P$, but it then fixes the positions of all the other $N-1$ electrons via the {\em co-motion} functions $\fv_i(\rv)$ \cite{SeiGorSav-PRA-07}. The strong-interaction limit of DFT is thus the generalization of the more familiar Wigner-crystal state to smooth densities. In the Wigner crystal state, in fact, the constraint of having a given density is relaxed, and  $\rho(\rv)$ becomes typically proportional to $\sum_i\delta(\rv-\rv_i^{(0)})$, losing any resemblence with the quantum mechanical $\lambda=1$ density of atoms and molecules.

From the condition that $E\pot$ have its minimum over the entire subspace $M$ and that the electrons be indistinguishable, one finds that the co-motion functions $\fv_i(\rv)$ must satisfy special properties, which are reported in \cite{SeiGorSav-PRA-07}. To determine the co-motion functions from the density $\rho(\rv)$, we use the observation \cite{SeiGorSav-PRA-07} that, since the position of the first electron determines the positions of all the others, the probability of finding the first electron in the volume element $d\rv$ around the position $\rv$ must be the same of finding the $i^{\rm th}$ electron in the volume element
$d\fv_i(\rv)$ around the position $\fv_i(\rv)$. This means that all 
the co-motion functions $\fv_i(\rv)$ must satisfy the differential equation
\beq
\rho(\fv_i(\rv)) d\fv_i(\rv)=\rho(\rv) d\rv,\qquad i=2,...,N.
\label{eq_detailedbalance}
\eeq
In order to construct the co-motion functions we thus have to find
the initial conditions for the integration of (\ref{eq_detailedbalance}) which also satisfy the properties reported in \cite{SeiGorSav-PRA-07}. An example of such calculations for spherical densities is carried out in \cite{SeiGorSav-PRA-07}. The strong interaction limit of DFT is thus entirely determined by the co-motion functions $\fv_i(\rv)$, which can be constructed from the density via Eqs.~(\ref{eq_detailedbalance}).

To obtain the intracule density $I^{\lambda\to\infty}(r_{12})$ corresponding to the distribution $|\Psi^{\lambda\to\infty}|^2$, we have to consider that the electron-electron distance only depends on the position of the first electron, $\rv$. By defining the $N(N-1)/2$ distances $d_{ij}(\rv)$ for which $|\Psi^{\lambda\to\infty}|^2$ is non zero, 
\beq
d_{ij}(\rv)=|\fv_i(\rv)-\fv_j(\rv)|,\qquad i,j=1,\dots,N,\quad i<
\label{eq_dij}
\eeq  
(with $\fv_1(\rv)\equiv \rv$), and by considering that each position $\rv$ has a probability $\frac{1}{N}\rho(\rv)$ \cite{SeiGorSav-PRA-07}, we obtain
\beq
4\pi\, r_{12}^2 \,I^{\lambda\to \infty}(r_{12})=\sum_{i>j=1}^N\int d\rv \,\frac{\rho(\rv)}{N}\,\delta(r_{12}-d_{ij}(\rv)).
\label{eq_Iinfty}
\eeq

\section{Application to atoms}
\label{sec_app}
We consider here the case of spherical densities, and we apply Eq.~(\ref{eq_Iinfty}) to few atoms. 
When $\rho(\rv)=\rho(r)$, the $\lambda\to\infty$ problem can be separated into an angular part and a radial part \cite{SeiGorSav-PRA-07}. The distance $r$ from the nucleus of one of the electrons can be freely chosen, and it then determines the distances from the nucleus of all the other $N-1$ electrons via radial co-motion functions $f_i(r)$, as well as all the relative angles $\alpha_{ij}(r)$ between the electrons \cite{SeiGorSav-PRA-07}. The radial co-motion functions $f_i(r)$ can be constructed as follows \cite{SeiGorSav-PRA-07}. Define an integer index $k$ running for odd $N$ from 1 to $(N-1)/2$, 
and for even $N$ from 1 to $(N-2)/2$. Then
\begin{eqnarray}
 f_{2k}(r)=\left\{
\begin{array}{lr}
 \nf^{-1}(2k-\nf(r)) & r\leq a_{2 k} \\
  \nf^{-1}(\nf(r)-2k) & r> a_{2k}
\end{array}
\right. \nonumber
 \\
 f_{2k+1}(r)=\left\{
\begin{array}{lr}
 \nf^{-1}(\nf(r)+2k) & r\leq a_{N-2 k} \\
  \nf^{-1}(2N-2k-\nf(r)) & r> a_{N-2k},
\end{array}
\right.
\label{eq_fradial}
\end{eqnarray}
where  $a_i=\nf^{-1}(i)$,  
\beq
\nf(r)=\int_0^r 4 \pi \,x^2 \rho(x)\,dx,
\eeq
and $\nf^{-1}(y)$ is the inverse function of $\nf(r)$.
For odd $N$, these equations give all the needed $N-1$ radial co-motion 
functions, while for even $N$ we have to add the last function,
\beq
f_N(r)= \nf^{-1}(N-\nf(r)).
\label{eq_fradialN}
\eeq
The relative angles $\alpha_{ij}(r)$ between the electrons can be found by minimizing numerically the electron-electron repulsion energy $\sum_{i>j} (f_i(r)^2+f_j(r)^2-2f_i(r)f_j(r) \cos\alpha_{ij})^{-1/2}$. The radial co-motion functions of Eqs.~(\ref{eq_fradial})-(\ref{eq_fradialN}) satisfy Eq.~(\ref{eq_detailedbalance}) for spherically symmetric $\rho$, 
\beq
4 \pi \,f_i(r)^2 \rho(f_i(r))\,|f'_i(r)|\,dr=4 \pi \,r^2 \rho(r)\,dr,
\label{eq_detailedbalancespher}
\eeq 
and, together with the minimizing angles $\alpha_{ij}(r)$, yield the minimum of $E\pot$ of Eq.~(\ref{eq_Epot}) for spherically symmetric $v(r)$ \cite{SeiGorSav-PRA-07}. Physically, the solution of Eqs.~(\ref{eq_fradial})-(\ref{eq_fradialN}) makes the $N$ electrons always be in $N$ different spherical shells, each of which contains, on average in the quantum mechanical problem (at $\lambda=1$), one electron. In the $\lambda\to\infty$ limit, the electrons become strictly correlated, and all fluctuations are suppressed (see, e.g., \cite{ZieTaoSeiPer-IJQC-00}): the space is divided into $N$ regions, each of which always contains exactly one electron.

For spherically symmetric densities, the electron-electron distances of Eq.~(\ref{eq_dij}) then become
\beq
d_{ij}(r)=\sqrt{f_i(r)^2+f_j(r)^2-2f_i(r)f_j(r) \cos\alpha_{ij}(r)}.
\label{eq_dijspher}
\eeq
We can invert Eq.~(\ref{eq_dijspher}) (if $d_{ij}(r)$ is non monotonous we have to invert each monotonous branch separately), and simplify Eq.~(\ref{eq_Iinfty}) into
\beq
4\pi\, r_{12}^2 \,I^{\lambda\to\infty}(r_{12})=\sum_{i>j=1}^N p(d_{ij}^{-1}(r_{12}))\,|d_{ij}'(d_{ij}^{-1}(r_{12}))|^{-1},\qquad p(r)=\frac{4\pi r^2 \rho(r)}{N}.
\label{eq_Ispherical}
\eeq
\begin{figure}
  \begin{center}
   \includegraphics{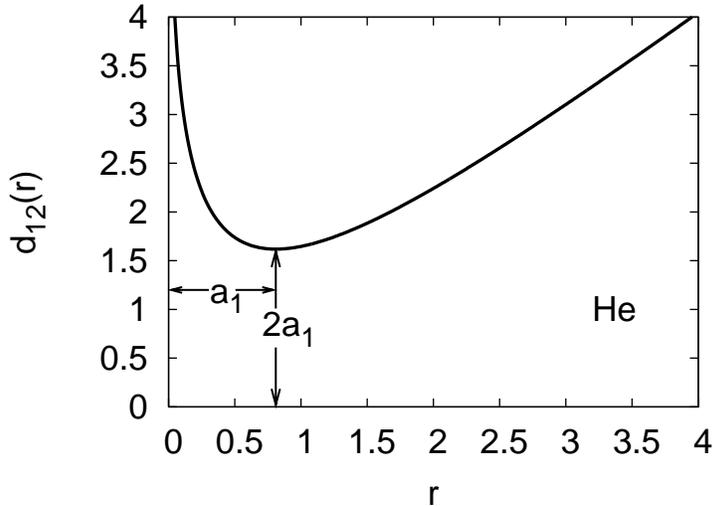}
   \caption{The electron-electron distance $d_{12}$ for the He atom density in the strong-interaction limit of DFT. $d_{12}$ is, in this limit, completely determined by the distance $r$ from the nucleus of one of the two electrons, $d_{12}=d_{12}(r)$. The value $r=a_1$, for which $d_{12}(r)$ has its minimum, corresponds to the radius of the sphere containing, on average in the quantum mechanical problem, one electron, $\int_0^{a_1} 4\pi\, r^2\,\rho(r) \,dr=1$. All quantities in Hartree atomic units.}
\label{fig_d12He}
  \end{center}
\end{figure}
\begin{figure}
  \begin{center}
   \includegraphics{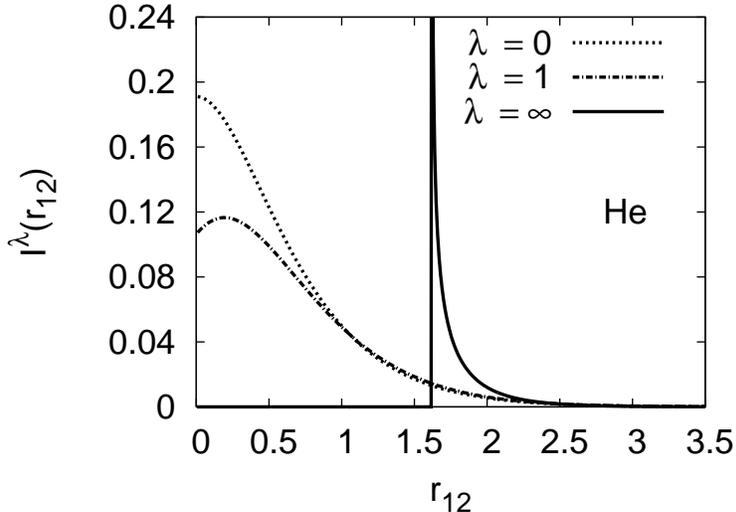}
   \caption{The intracule density of the He atom along the linear adiabatic connection of DFT: the three intracules correspond to three systems with the same one-electron density $\rho(\rv)$ and electron-electron interaction $\lambda/r_{12}$. The intracules at $\lambda=0$ (KS system) and $\lambda=1$ (physical system) have been constructed from the accurate variational wavefunction of Ref.~\cite{FreHuxMor-PRA-84} (see also \cite{UmrGon-PRA-94} and \cite{GorSav-PRA-05}). The intracule at $\lambda=\infty$ is obtained in this work, as described in the text. All quantities in Hartree atomic units.}
\label{fig_IHe}
  \end{center}
\end{figure}

As a starting point, it is instructive to analyze the simple case of the He atom. Here, we have used the accurate variational wavefunction of Ref.~\cite{FreHuxMor-PRA-84} (see also \cite{UmrGon-PRA-94} and \cite{GorSav-PRA-05}) to generate the density $\rho(\rv)$, from which $I^{\lambda\to\infty}(r_{12})$ is constructed. From the same accurate variational wavefunction, we have also computed the intracules of the KS system (see \cite{GorSav-PRA-05}), $I^{\lambda=0}(r_{12})$, and of the physical system, $I^{\lambda=1}(r_{12})$, which will be compared to $I^{\lambda\to\infty}(r_{12})$. When $N=2$, in the $\lambda\to\infty$ limit the relative angle between the electrons becomes always $\alpha_{12}=\pi$ (maximum angular correlation), and we have only one co-motion function, $f_2(r)$, which fully determines the electron-electron distance \cite{Sei-PRA-99,SeiGorSav-PRA-07},
\beq
f_2(r)=\nf^{-1}(2-\nf(r)),\qquad d_{12}(r)=r+f_2(r).
\eeq
The function $d_{12}(r)$ is reported in Fig.~\ref{fig_d12He}. It has a minimum for $r=a_1=\nf^{-1}(1)$, the radius of the shell containing, on average in the quantum mechanical problem, one electron. We have $f_2(a_1)=a_1$, and $d_{12}(a_1)=2a_1$. The two electrons, thus, never get closer than $2 a_1$, so that $I^{\lambda\to\infty}(r_{12}<2a_1)=0$. From the indistinguishability of the two electrons, we have the property \cite{Sei-PRA-99,SeiGorSav-PRA-07} $f_2(f_2(r))=r$, which, when combined with Eq.~(\ref{eq_detailedbalancespher}), shows that the two invertible branches of $d_{12}(r)$ (corresponding to $0\le r<a_1$ and to $r>a_1$) give the same contribution to Eq.~(\ref{eq_Ispherical}). We can thus just invert the function $d_{12}(r)$ in $r\in[0,a_1]$ and multiply the result by 2. This is a general property, also valid for $N>2$ \cite{SeiGorSav-PRA-07}: we can always invert the functions $d_{ij}(r)$ in the domain $r\in[0,a_1]$, and then multiply the result by $N$. In Fig.~\ref{fig_IHe} we report the intracule densities of the He atom for $\lambda=0$ (KS), $\lambda=1$ (physical) and $\lambda\to\infty$. Because $d_{12}'(a_1)=0$, $I^{\lambda\to\infty}(r_{12})$ has an integrable divergence when $r_{12}\to 2 a_1^+$,
\beq
4\pi\, r_{12}^2\,I^{\lambda\to\infty}(r_{12})\Big|_{r_{12}\to 2 a_{1}^+}\to
\frac{p(a_1)^{3/2}}{\sqrt{-p'(a_1)(r_{12}-2 a_1)}},
\qquad  p(r)=2\pi r^2\rho(r).
\eeq
Divergences come from the fact that the quantities we calculate here are the distributions towards which the physical quantities tend when $\lambda\to\infty$. This aspect is clarified with a simple example in the Appendix of Ref.~\cite{SeiGorSav-PRA-07}.
\begin{figure}
  \begin{center}
   \includegraphics{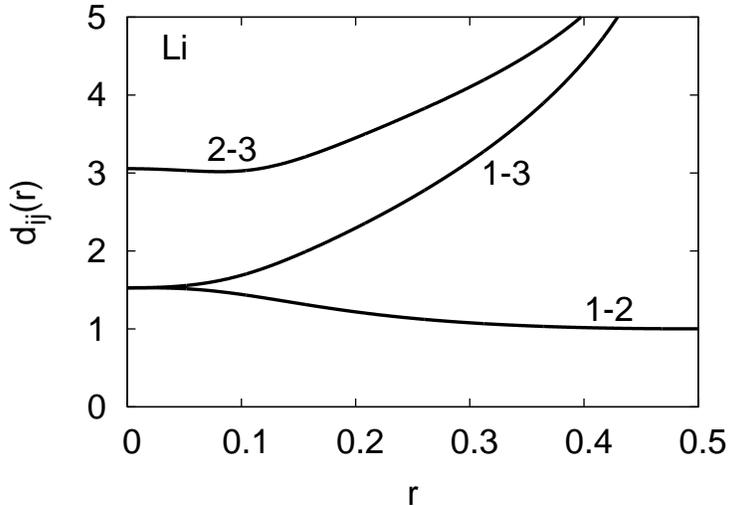}
   \caption{The electron-electron distances $d_{ij}$ for the Li atom density in the strong-interaction limit of DFT. The distances $d_{ij}$ are, in this limit, completely determined by the distance $r$ from the nucleus of one of the electrons, $d_{ij}=d_{ij}(r)$. The value of $r$ is varied here between 0 and $a_1$, the radius of the sphere containing, on average in the quantum mechanical problem, one electron, $\int_0^{a_1} 4\pi\, r^2\,\rho(r) \,dr=1$. As explained in the text, the case $r>a_1$ does not need to be considered, since it simply corresponds to interchanging two or more electrons, yielding the same values for the electron-electron distances. All quantities in Hartree atomic units.}
\label{fig_distLi}
  \end{center}
\end{figure}
\begin{figure}
  \begin{center}
   \includegraphics{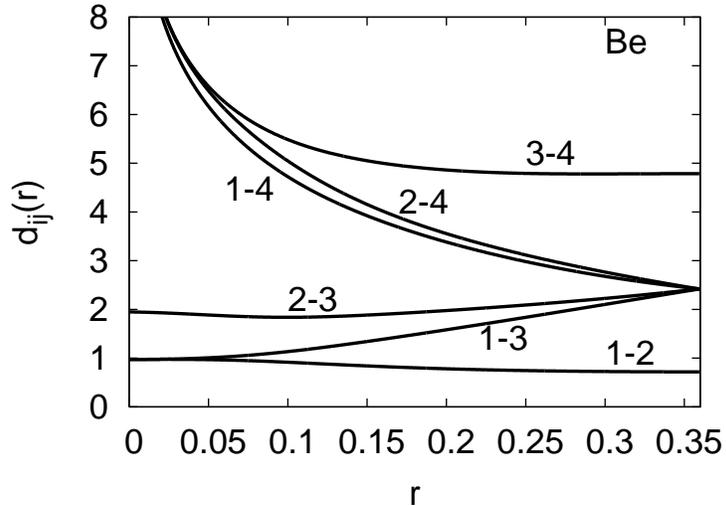}
   \caption{Same as Fig.~\ref{fig_distLi} for the Be atom density.}
\label{fig_distBe}
  \end{center}
\end{figure}

Using Eqs.~(\ref{eq_fradial})-(\ref{eq_fradialN}), we have calculated the electron-electron distances $d_{ij}(r)$ of Eq.~(\ref{eq_dijspher}) for Li, Be and Ne. For the Li atom, we have used the fully correlated density of Bunge \cite{Bun-PRIV-XX}, and for the Be and the Ne atoms the accurate densities of Ref.~\cite{FilGonUmr-INC-96}. All calculations are done numerically, on a grid. In Figs.~\ref{fig_distLi} and \ref{fig_distBe}, we report $d_{ij}(r)$ for Li and Be when $0\le r\le a_1$. Electrons are labeled with numbers 1, 2, 3,..., meaning that electron 1 is in the shell $0\le r\le a_1$, electron 2 is in the shell $a_1\le f_2(r) \le a_2$, and so on. Correspondingly, the distances are labeled 1-2, 1-3, etc. As discussed for the case of the He atom (for further details see \cite{SeiGorSav-PRA-07}), we only need to consider $N$ times this situation, since exchanging two or more electrons always correspond to the same physics, and thus to the same values of the electron-electron distances.
\begin{figure}
  \begin{center}
   \includegraphics{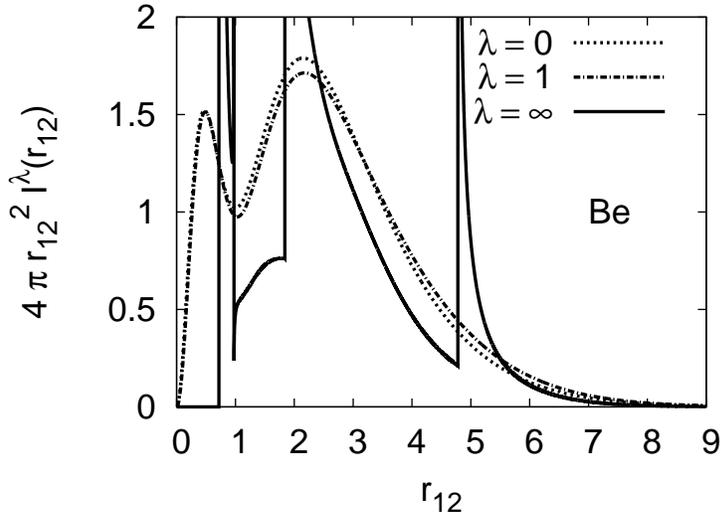}
   \caption{The intracule density of the Be atom multiplied by the volume element $4\pi r_{12}^2$ along the linear adiabatic connection of DFT: the three intracules correspond to three systems with the same one-electron density $\rho(\rv)$ and electron-electron interaction $\lambda/r_{12}$. The intracule at $\lambda=0$ (KS system) has been obtained from the accurate KS potential of Ref.~\cite{FilGonUmr-INC-96} and  the intracule at  $\lambda=1$ (physical system) is taken from Ref.~\cite{TouAssUmr-JCP-07}. The intracule at $\lambda=\infty$ is calculated in this work, as described in the text. All quantities in Hartree atomic units.}
\label{fig_IBe}
  \end{center}
\end{figure}
\begin{figure}
  \begin{center}
   \includegraphics{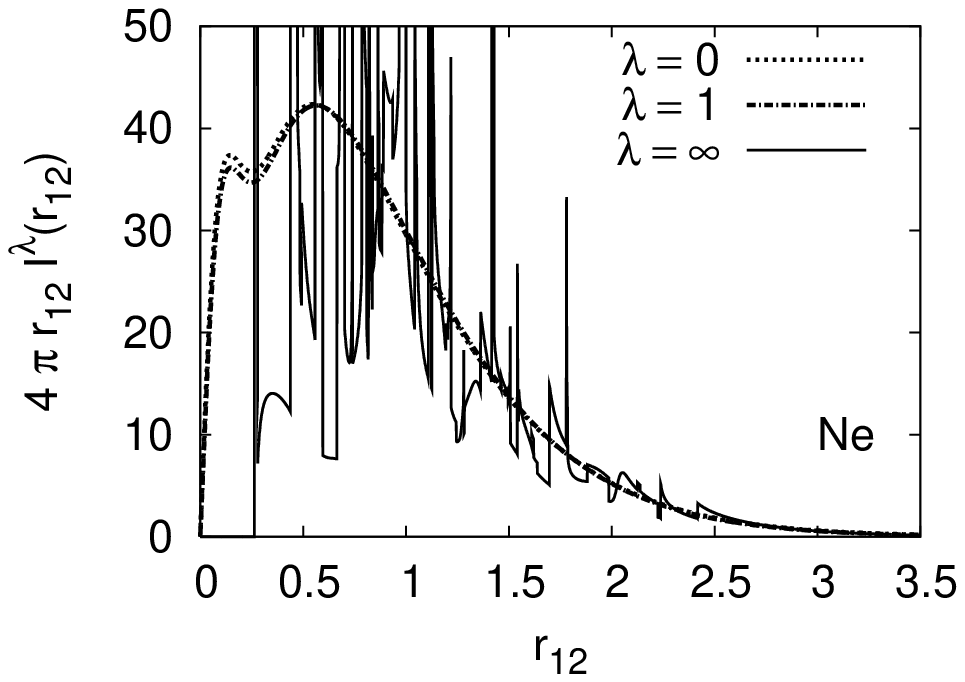}
   \caption{The intracule density of the Ne atom multiplied by $4\pi r_{12}$ along the linear adiabatic connection of DFT: the three intracules correspond to three systems with the same one-electron density $\rho(\rv)$ and electron-electron interaction $\lambda/r_{12}$. The intracule at $\lambda=0$ (KS system) has been obtained from the accurate KS potential of Ref.~\cite{FilGonUmr-INC-96} and  the intracule at  $\lambda=1$ (physical system) is taken from Ref.~\cite{TouAssUmr-JCP-07}. The intracule at $\lambda=\infty$ is calculated in this work, as described in the text. The area under each curve gives the expectation $\langle\Psi^\lambda|\hat{V}_{ee}|\Psi^\lambda\rangle$. All quantities in Hartree atomic units.}
\label{fig_INe}
  \end{center}
\end{figure}

In Fig.~\ref{fig_IBe} we show the intracule density multiplied by the volume element $4\pi r_{12}^2$ at $\lambda=0$, $\lambda=1$ and $\lambda\to\infty$ for the Be atom. The case of the Ne atom is displayed in Fig.~\ref{fig_INe}, where we show the intracule densities multiplied by $4\pi\,r_{12}$: the area under each curve gives the expectation $\langle\Psi^\lambda|\hat{V}_{ee}|\Psi^\lambda\rangle$.
In both cases, the KS intracules at $\lambda=0$ have been constructed from the accurate Kohn-Sham potentials of Ref.~\cite{FilGonUmr-INC-96}, while the intracules of the physical system ($\lambda=1$) are obtained from variational Quantum Monte Carlo results \cite{TouAssUmr-JCP-07}. Integrable divergences in the $\lambda\to\infty$ intracules appear, as for the He atom case, at elecron-electron distances for which $d_{ij}'(r)=0$. Notice that, due to the strict correlation at $\lambda\to\infty$, there is a finite minimum distance $r_{ij}^{\rm min}>0$ between any pair of electrons, so that $I^{\lambda\to\infty}(r_{12})= 0$ when $r_{12}$ is less than the smallest of the $r_{ij}^{\rm min}$.

\section{Discussion of Results}
\label{sec_disc}
The $\lambda\to\infty$ limit of DFT describes the case of maximal angular correlation and maximal radial correlation between the electrons, compatible with the constraint that the probability of finding one electron at postion $\rv$ be equal to $\rho(\rv) d\rv$, where the density $\rho(\rv)$ corresponds to the quantum mechanical hamiltonian of Eq.~(\ref{eq_Hphys}). For atoms, we see from Figs.~\ref{fig_IHe}, \ref{fig_IBe} and \ref{fig_INe} that the intracule of the physical system ($\lambda=1$) is much closer to the KS intracule ($\lambda=0$) than to the $\lambda\to\infty$ limit, as expected for weakly correlated systems. In more correlated situations like streched bonds, we can expect the $\lambda=1$ case to be more in between the $\lambda=0$ and the $\lambda\to\infty$ limits. The investigation of such cases will be the object of future work.
 
What can we learn from the intracule densities in the $\lambda\to\infty$ limit?
As a first step, we have reconsidered the results of Ref.~\cite{GorSav-PRA-05} for the He-like ions. In that work, we had shown that a good approximation for $w_{c}^\lambda$ of Eq.~(\ref{eq_wc}) was given by the screening potential of a sphere of uniform density, charge 1, and radius $\overline{r}_s$ to be determined,
\beq
w_c^\lambda(r_{12})\approx -
\left(\frac{4\pi}{3} \overline{r}_s^3\right)^{-1}\int_{|\rv|\le \overline{r}_s} 
  \frac{\lambda}{|\rv - \rv_{12}|}\,d \rv.
\label{eq_wcOv}
\eeq 
This approximation was first introduced by Overhauser for the uniform electron gas \cite{Ove-CJP-95}, where $\overline{r}_s$ was set equal to the usual density parameter $r_s$, i.e., the radius of the sphere containing, on average, one electron. In Ref.~\cite{GorPer-PRB-01} it was shown, by comparison with Quantum Monte Carlo results,  that for the uniform electron gas the Overhauser potential very accurately recovers the short-range part ($r_{12}\le r_s$) of $I(r_{12})$. For the He-like ions, setting \cite{GorSav-PRA-05} $\overline{r}_s^3=\frac{3}{4\pi \overline{\rho}}$, where $\overline{\rho}$ is an average density, $\overline{\rho}=\frac{1}{N}\int d\rv \rho(\rv)^2$, also yields accurate results for the  short-range part of $I(r_{12})$, with ``on-top'' values $I(0)$ essentially indistinguishable from those coming from very accurate Hylleras-type variational wavefunctions. Here, we realized that the values $\overline{r}_s$ used in Ref.~\cite{GorSav-PRA-05} for the He-like ions are, within few percents, equal to the values $a_1$ of the radii of the sphere (centered in the nucleus) containing on average one electron, thus making Eq.~(\ref{eq_wcOv}) work equally well for the uniform electron gas and for the He series, with the same choice for the screening length $\overline{r}_s$. As shown in the previous section, the value $a_1$ plays a special role in the $\lambda\to\infty$ limit of the He-like ions. We can thus hope to learn from the $\lambda\to\infty$ intracules something about ``multiple screening lenghts'' for approximating $w_c^\lambda(r_{12})$ for many-electron systems of nonuniform density.
\begin{figure}
  \begin{center}
   \includegraphics{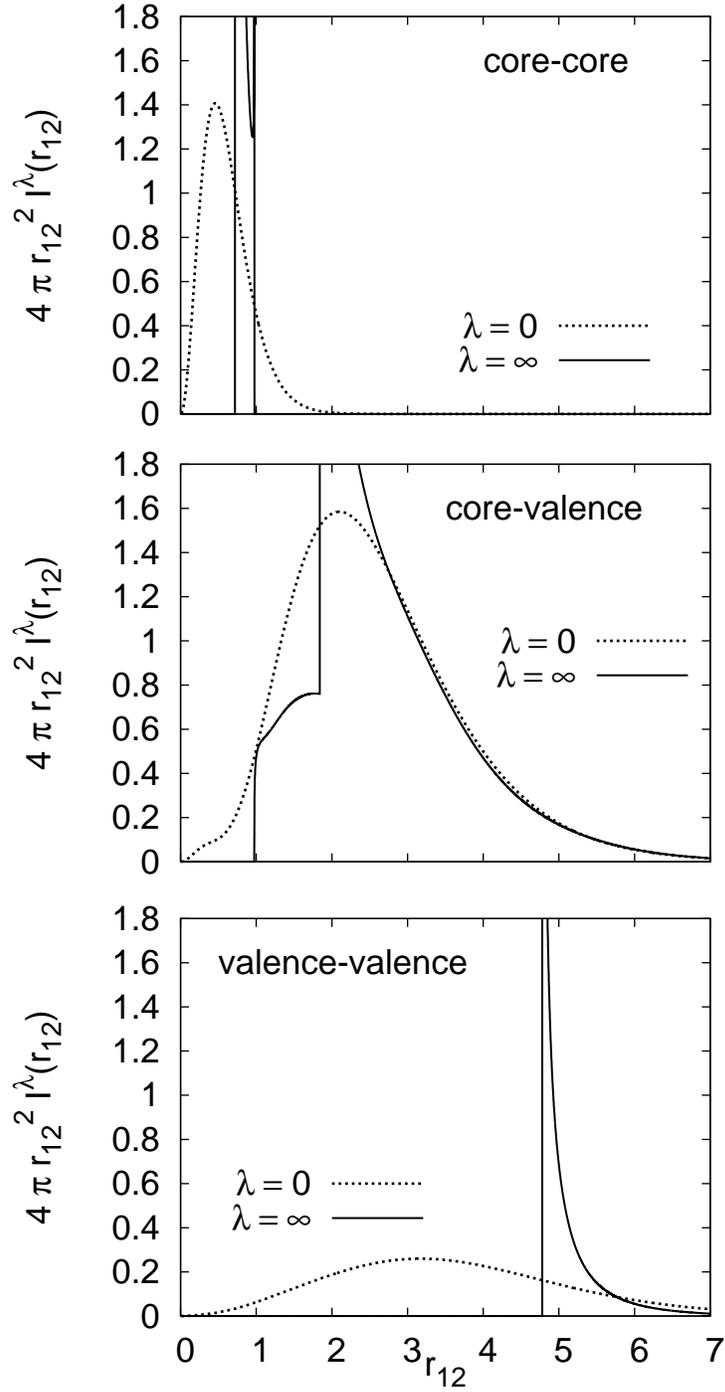}
   \caption{Be atom: the core-core, core-valence and valence-valence contributions to the intracule densities at $\lambda=0$ and $\lambda=\infty$ of Fig.~\ref{fig_IBe}.}
\label{fig_Becccvvv}
  \end{center}
\end{figure}
\begin{figure}
  \begin{center}
   \includegraphics{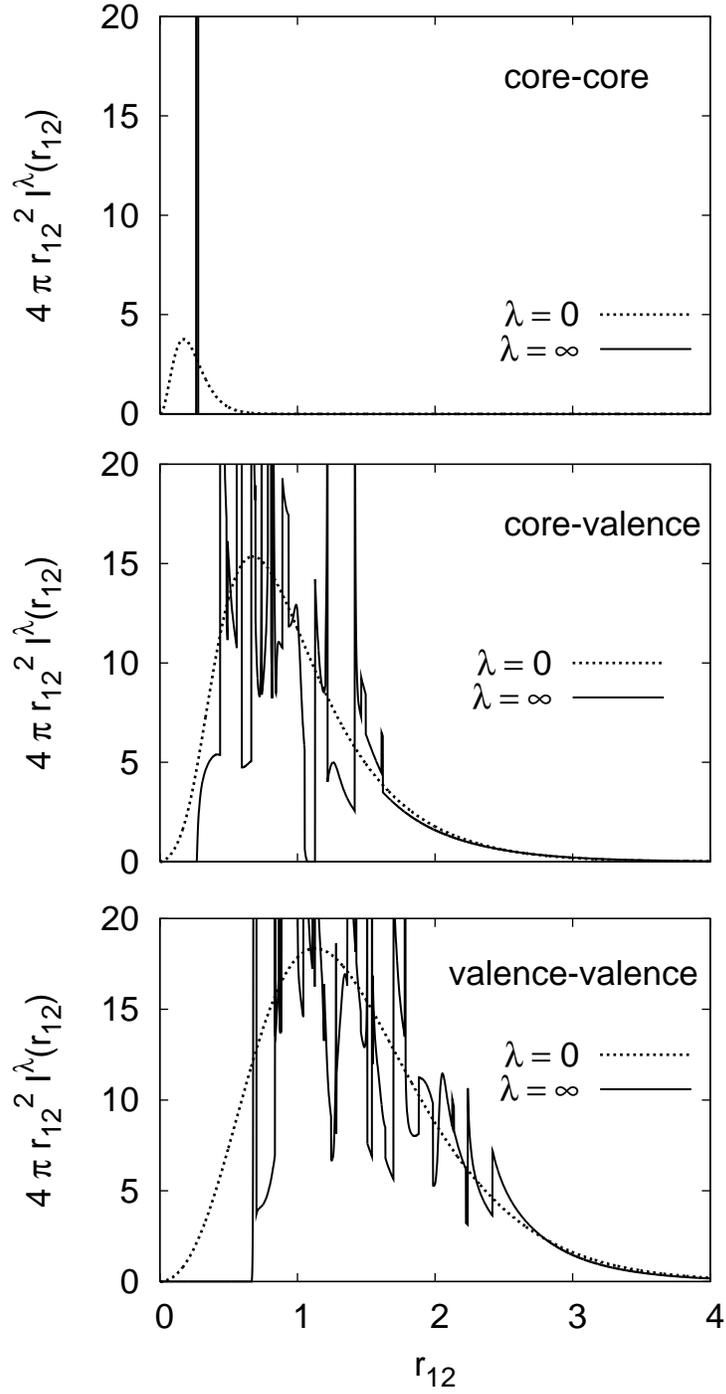}
   \caption{Ne atom: the core-core, core-valence and valence-valence contributions to the intracule densities at $\lambda=0$ and $\lambda=\infty$ of Fig.~\ref{fig_INe}.}
\label{fig_Necccvvv}
  \end{center}
\end{figure}

To this purpose, and with the idea in mind that to have a size-consistent method we need to use localized geminals in Eqs.~(\ref{eq_eqsforI}), we have further analyzed our results by dividing them into core-core, core-valence, and valence-valence contributions, comparing the $\lambda\to\infty$ and the $\lambda=0$ case. We consider here the Be and the Ne atoms. For the $\lambda\to\infty$ case, the core-core contribution comes from the distance 1-2, corresponding to the two electrons that are in the sphere containing, on average in the quantum mechanical problem, 2 electrons. The distances of electrons 1 and 2 from the other electrons define the core-valence contribution, and the rest is the valence-valence part. For the KS system ($\lambda=0$), we have orbitals, so the three contributions are defined in the usual way, using the quantum mechanical shells (1$s^2$ for core-core, etc.). The three contibutions are shown in Fig.~\ref{fig_Becccvvv} for Be and in Fig.~\ref{fig_Necccvvv} for Ne. In the core-valence case of Be, we see that the extremely correlated $\lambda\to\infty$ limit differs from its KS counterpart only in the short-range part, $r_{12}\lesssim 2.5$. The valence-valence case of Be resembles the two-electron case of the He atom. The core-valence and the valence-valence contributions to the Ne atom also show, essentially, correlation of short-range type, even in the extreme $\lambda\to\infty$ case. In our future work, we plan to use these results to build and test approximations for $w_c^\lambda(r_{12})$.

\section{Conclusions and perspectives}
\label{sec_conc}
We have calculated, for the first time, the intracule densities for small atoms in the strong-interaction limit of density functional theory. Our results can be useful to better understand correlation in the density functional theory framework, and to build approximations for correlation energy functionals based on intracules. Our future work on this subject will address several points:
\begin{enumerate}
 \item The generalization of this calculation to non-spherical densities;
\item The study of the next leading term in the $\lambda\to\infty$ limit, thus including zero-point-motion oscillations;
\item The use of these results to fully develop the ideas of Refs.~\cite{GorSav-PRA-05,GorSav-PM-06,Nag-JCP-06,GorSav-IJMPB-07,GorSav-JCTC-07}, in which an approximation for the correlation energy in density functional theory is contructed from effective equations for the intracule density.  

\end{enumerate}

\section*{Acknowledgments}
We thank Carlos Bunge for the fully correlated density of the Li atom, Cyrus Umrigar for the accurate densities and Kohn-Sham potentials of Be and Ne, and Julien Toulouse for the QMC intracule densities of Be and Ne.



\end{document}